\documentclass[journal]{IEEEtran}

\usepackage[utf8]{inputenc} 

\usepackage{graphicx} 

\usepackage{booktabs} 
\usepackage{array} 
\usepackage{paralist} 
\usepackage{verbatim} 
\usepackage{subfig} 
\usepackage{amsmath}
\usepackage{amsfonts}
\usepackage{amsthm}
\usepackage{cite}
\usepackage{balance}
\usepackage[ruled,lined,commentsnumbered, linesnumbered]{algorithm2e}
\usepackage{graphicx}
\usepackage{color}
\usepackage{multirow}
\usepackage[table,xcdraw]{xcolor}
\usepackage{enumerate}

\graphicspath{{./Figures/}}

\def\0{\mathbf{0}}

\def\v{\mathbf{v}}
\def\w{\mathbf{w}}
\def\x{\mathbf{x}}

\def\z{\mathbf{z}}

\def\A{\mathbf{A}}

\def\X{\mathbf{X}}
\def\Y{\mathbf{Y}}

\def\-{\mbox{-}}

\renewcommand{\baselinestretch}{1.0}

\begin{document}

\title{Improving Observability of Stochastic Complex Networks under the Supervision of \\ Cognitive Dynamic Systems}

\author{Mehdi~Fatemi,~\IEEEmembership{Member,~IEEE},
	    Peyman~Setoodeh,~\IEEEmembership{Member,~IEEE,}
            Simon~Haykin,~\IEEEmembership{Life~Fellow,~IEEE} 

\thanks{This work was supported by the National Science and Engineering Research Council (NSERC)
of Canada.}
\thanks{M. Fatemi and S. Haykin are with the Cognitive Systems Laboratory, McMaster University, Hamilton, Ontario, Canada (e-mail: {mehdi.fatemi@ieee.org}, {haykin@mcmaster.ca}).}
\thanks{P. Setoodeh is with the School of Electrical and Computer Engineering, Shiraz University, Shiraz, Iran (e-mail: {setoodeh@ieee.org}).}
}

\mathchardef\mhyphen="2D

\maketitle

\begin{abstract}
Much has been said about observability in system theory and control; however, it has been recently that observability in complex networks has seriously attracted the attention of researchers. This paper examines the state-of-the-art and discusses some issues raised due to ``complexity'' and ``stochasticity''. These unresolved issues call for a new practical methodology. For stochastic systems, a degree of observability may be defined and the observability problem is not a binary (i.e., yes-no) question anymore. Here, we propose to employ a goal-seeking system to play a supervisory role in the network. Hence, improving the degree of observability would be a valid objective for the supervisory system. Towards this goal, the supervisor dynamically optimizes the observation process by reconfiguring the sensory parts in the network. A cognitive dynamic system is suggested as a proper choice for the supervisory system. In this framework, the network itself is viewed as the environment with which the cognitive dynamic system interacts. Computer experiments confirm the potential of the proposed approach for addressing some of the issues raised in networks due to complexity and stochasticity.
\end{abstract}


%

\section{Introduction}

In 1977, Herbert Simon wrote \cite{simon1977}, p. 258:
\begin{quotation}
\noindent ``To a Platonic mind, everything in the world is connected to everything else---and perhaps it is. Everything is connected but some things are more connected than others.''
\end{quotation}
The point he is emphasizing is \emph{connectivity}, which is at the heart of complex networks. Indeed, the complexity of networks manifests itself in how dense and with what kind of structure, the edges are distributed in a network with arbitrary large number of nodes.

In the realm of network science, an extremely important issue to be addressed in many of real-world
applications is how to acquire sufficient information about a network with
minimal computational effort. In other words, the problem of interest is to understand why a complex network with high connectivity behaves in a certain way by accessing only a small subset of nodes. This problem is subsumed under the broader problem of network \emph{observability}. Knowing whether or not a network is observable would be critical because in large networks, it is often impractical or even impossible to monitor all nodes' states. On the other hand,
in many real-world applications, all the states are not necessarily accessible to the outside world. Therefore, there is a need to reconstruct (i.e., estimate) those states on the sole basis of observing other variables, which are related to those states as well as accessible for measurements.


Although it is a classic and well-known problem in system theory and control, observability in network science is relatively new, mostly started with the prominent work reported in \cite{Liu2013_Observability}. For deterministic networks, the proposed algorithm in \cite{Liu2013_Observability} yields the minimum number of nodes in the network that should be monitored in order to satisfy the requirement for observability. It also provides subsets of nodes from which the monitor nodes should be selected. However, extending the results to stochastic networks aiming at estimating the state of the network does not seem to be that straightforward. As a matter of fact, having the proposed framework of \cite{Liu2013_Observability} in mind, some of the simulation results obtained for complex stochastic networks (especially those with dense structures) may seem counterintuitive. Hence, for estimating a network's state in face of model uncertainties and imperfect measurements, additional steps must be taken. We distinguish our paper from \cite{Liu2013_Observability} in two accounts: a) complexity in terms of edge density, and b) stochasticity. 

Here, we propose to implement a \textit{controlled-sensing} mechanism in the network. By taking this approach, a supervisory system would be responsible for \textit{reconfiguration} of the sensory parts in the network in order to dynamically \textit{optimize the observation} process. A \textit{cognitive dynamic system} (CDS) in the sense described in \cite{haykin_CDS} will be able to perfectly play the role of the supervisory system, where the stochastic network of interest is viewed as the environment with which the CDS interacts. A CDS is built on Fuster's paradigm of cognition, which suggests five pillars for a cognitive system: perception-action cycle, memory, attention, intelligence, and language \cite{fuster_cortex_mind}. Perceptual and executive parts of the perception-action cycle as well as memory are physical entities, attention is algorithmic, and intelligence emerges due to the interactions among the former three pillars. Language will play a key role, when we have a network of cognitive systems.

Following this new way of thinking, the CDS, which acts as a supervisor over a given network of
interest, tries to reconstruct the hidden states of the network based on the information it gathers from monitor nodes (i.e., a selected subset of nodes whose outputs are accessible to the CDS). The perceptual part of the CDS employs Bayesian filtering for reconstruction of entire state of the network. Furthermore, through the use of \textit{cognitive control} \cite{Haykin2012_CC, Fatemi2014}, the executive part of the CDS tries to improve accuracy of the reconstructed state from each global cycle of perception-action to the next. To this end, a quantitative measure for the lack of information in the state posterior is also computed in the cognitive perceptor, which is passed on to the cognitive controller as the feedback information.
The cognitive controller will then use this information to reconfigure the sensory parts of the network by \textit{rearranging} the monitor nodes in such a way that the available information to the perceptor is maximized in the following cycles. In addition to rearrangement of monitors, the CDS may also have to increase the number of monitor nodes or remove redundant ones (i.e., nodes with minimal contribution in acquiring information).

In the proposed approach, learning and planning stages involved in cognitive control provide enough flexibility to handle different situations that may occur in the network of interest. In this regard, a few points are worth mentioning:
\begin{itemize}
\item Cognitive control can directly incorporate any practical constraints such as limitation on the number of nodes that can be monitored (i.e., number of deployed sensors).
\item Due to design parameters such as learning and discount factors as well as size and depth of the planning stage, the methodology can be adapted for different practical applications.
\item The required computations for implementing the proposed methodology can be performed either online or partially offline:
\begin{enumerate}[i)]
\item In the online implementation, cognitive controller and Bayesian filter find the best set of monitor nodes taking some prescribed constraints into account. Moreover, the selection process happens in a cyclic manner from each global cycle of perception-action to the next.
\item In the partially offline implementation, the proposed methodology is used as the basis for Monte Carlo simulations, which provide clues about the best set of monitor nodes considering the practical constrains. Here, the preferred sets of monitor nodes for different working conditions are found and stored beforehand. This way, an appropriate set of monitor nodes for current operational conditions will be recalled from the stored data and therefore the amount of computation that must be performed on the fly will significantly decrease.
\end{enumerate}
\end{itemize}

The rest of the paper is organized as follows: Section \ref{sec_net_sci} reviews some basic concepts from network science as the required background for the following sections. Next, in Section \ref{sec_obs}, the problem of stochastic observability is discussed in detail with emphasis on network observability. Section \ref{sec_pac} explains how complex networks can be viewed as the environment with which a cognitive dynamic system interacts. This way, the CDS plays the role of a supervisor that is responsible for improving network observability. Advantages of the proposed approach are shown through a set of computationl experiments in Section \ref{experiments} for both linear and nonlinear case studies. Finally, Section \ref{sec_discussion} concludes the paper by highlighting the key results and drawing lines for future research.

\section{Brief Account on Network Science} \label{sec_net_sci}

Regarding the critical role that networks play in shaping and sustaining our modern societies, the study of \textit{complex networks} has been expanding across diverse scientific disciplines over the last two decades \cite{Cohen2010}. This relatively new branch of science has began to be referred to as \emph{network science}.

\subsection{Networks with Stochastic Dynamics} \label{sec_network}

   A number of entities that have interactions with each other (i.e., linked in a physical and/or mathematical sense) form a \emph{network}. The underlaying \emph{topology} of a network is mathematically described by a graph, where each node represents one entity and edges show the interactions between the nodes they connect. Moreover, each entity in the network (i.e., each node in the graph) attributes to a \emph{state}. In reality, each state is a realization of a physical quantity, such as the electrical load on a power station, the density of a chemical compound in a biomedical receiver, or the amount of an item in a warehouse.

In order to be mathematically precise, the following set of definitions are recalled from graph theory \cite{Cohen2010}:


\noindent \textbf{Definition (digraph)}. A \textit{digraph} (directed graph) $G(N,L)$ is determined by a pair of sets:
  \begin{enumerate}
    \item A set of nodes, $N$ with $|N|=n$, where $n$ is called the graph size.
    \item A set of directed edges:
   \end{enumerate}
        \begin{align}
        \nonumber L = \{(i,j) \mbox{ iff there exists an edge from $i$ to $j$ for } i,j\in N \}.
        \end{align}


\noindent \textbf{Definition (incident matrix)}. The \textit{weighted incident matrix} (or simply incident matrix) of digraph $G(N,L)$ is a square matrix, $\A\in \mathbb{R}^{n\times n}$, which has a row and a column for each node. If there is a link from node $i$ to node $j$ in the digraph, the corresponding element of the incident matrix $\A_{ij}$, which represents the dependency \textit{weight} of node $j$ on $i$, will be nonzero. Otherwise, the entry $\A_{ij}$ will be zero. Hence, in general, the incident matrix of a digraph is asymmetric. If there are edges in the network that connect some nodes to themselves (i.e., if self-loops exist in the digraph), the corresponding diagonal elements of $A$ will be nonzero.


A network may represent a linear stochastic system that satisfies the Markovian assumption. In this case, the following pair of process and monitor equations provide a dynamic model for the network:
\begin{align}
\left\{\begin{array}{l}\mbox{Process equation: }~~~~\textbf{x}_{k+1} = \textbf{A}_{k}^{T}\textbf{x}_{k} + \textbf{v}_{k} \\
\mbox{Monitor equation: }~~~\textbf{z}_{k}=\textbf{C}_{k}\textbf{x}_{k} + \textbf{w}_{k}\end{array}\right.
\label{eq_state_space}
\end{align}
where $\textbf{A}_{k}$ is the incident matrix of the corresponding digraph $G$ at cycle $k$ that may vary in the course of time and $T$ denotes the transpositional operator. In this model, state vectors of the digraph's nodes, $\x_{k}^{(i)}$, are concatenated to form the augmented vector $\textbf{x}_{k}= \{\x_{k}^{(i)}\}_{i=1}^{n}$ that represents the whole network state. The evolution of the network state in the course of time is governed by the above process equation in which the process noise, $\v_k$, takes account of model uncertainties. It is assumed that a subset of nodes, $M\subseteq N$ with $|M|=m\leq n$, is available for monitoring, from which $q \leq m$ nodes are chosen as monitor nodes. Therefore, there are ${m \choose q} = \frac{m!}{q!(m-q)!}$ different options for choosing $q$ monitor nodes from $m$ accessible nodes. Similarly, the observed outputs of the monitored nodes, $\z_{k}^{(j)}$, are concatenated to form the augmented measurement vector $\textbf{z}_{k}= \{\z_{k}^{(j)}\}_{j=1}^{q}$. The above monitor equation~\footnote{In the control literature, the second equation in (\ref{eq_state_space}) is called measurement or output equation. In the network context, the measurements (i.e. observables) are provided by nodes that are chosen to be monitored (i.e. monitor nodes). Hence, the term monitor equation was adopted.} describes the relationship between the state and measurement vectors, where the measurement noise, $\w_k$, takes account of measurement uncertainties. Matrix $\textbf{C}_{k}$, has a row associated with the output of every monitor node at cycle $k$.

For the sake of brevity, in this paper, we assume that the random processes $\textbf{v}$ and $\textbf{w}$ are both zero-mean, white and mutually independent. Also, we solely focus on Gaussian environments, i.e., $v_{k} \sim \mathcal{N}(0, \textbf{Q}_{k})$ and $w_{k}\sim \mathcal{N}(0, \textbf{R}_{k})$, where $\textbf{Q}_{k}$ and $\textbf{R}_{k}$ denote the \emph{covariance} matrices of process and monitor noises, respectively. However, the application of cognitive control is not limited to Gaussian models.


The same modelling philosophy can be equally applied to stochastic nonlinear systems regarding the fact that for a nonlinear system, as discussed in \cite{Liu2013_Observability}, there exists a unique inference diagram (i.e., a digraph). In such cases, the network, which is mathematically described by the corresponding inference graph, represents the stochastic nonlinear system under study. The state-space model will then take the following form:
\begin{align}
\left\{\begin{array}{l}\mbox{Process equation: }~~~~\textbf{x}_{k+1} = \textbf{f}_{k}(\textbf{x}_{k}) + \textbf{v}_{k} \\
\mbox{Monitor equation: }~~~\textbf{z}_{k} = \textbf{g}_{k}(\textbf{x}_{k}) + \textbf{w}_{k}\end{array}\right.
\label{eq_state_space_nonlin}
\end{align}
If we have direct access to the states of nodes that are monitored, the above nonlinear monitor equation will be reduced to a linear one.

The linear and nonlinear state equations in (\ref{eq_state_space}) and (\ref{eq_state_space_nonlin}) are discrete-time models. The developed framework can be equally applied for continuous-time processes. However, in such cases, a hybrid (i.e., continuous-discrete) version of the Bayesian filter would be required for network state reconstruction.

Now that we have covered the relationship between state-space models, digraphs, and networks, we need to know how \emph{edge distribution} as well as \emph{edge density} affect the observability of complex networks if the number of nodes does not change. Answers to these questions will help for better sensor design in real-world applications. In order to set the stage for answering these key questions, we take a look at some well-known network topologies.

\subsection{Two Basic Network Topologies of Practical Importance:} Among different classes of networks, we consider Erd\H{o}s-R\'{e}nyi and scale-free random networks for their importance in modelling real-world networks \cite{jackson2008}:

\subsubsection{Erd\H{o}s-R\'{e}nyi (ER) Networks}

Named after P. Erd\H{o}s and A. R\'{e}nyi \cite{erdos1960}, in the growing process of this class of networks, a connection (an edge in the graph) may be produced between each pair of nodes with equal probability $p$, independent of the other edges. In their seminal paper, Erd\H{o}s and R\'{e}nyi provided a detailed behavioural analysis for such networks for different values of $p$. As a result, ER networks have become the most basic class in complex-network studies.

\subsubsection{Scale-free Networks}

A scale-free network is a network whose degree distribution follows a power law, at least asymptotically. To be more precise, let the fraction of nodes in the network that have $k$ connections to other nodes be denoted by $P(k)$. Then, for large values of $k$ we will have $P(k) \sim k^{-\gamma}$, where $\gamma$ is a parameter whose value is typically in the range $\gamma \in (2,3)$, although occasionally it may lie outside of this interval \cite{barabasi1999, choromanski2013}. Many of the real-world networks are thought to be somehow scale-free. Examples include social and collaborative networks, internet networks including the World Wide Web, some financial networks, protein-protein interaction networks, and airline networks.

Next section provides a formal definition of observability in the context of stochastic networks and suggests a way for improving the observability.

\section{Observability of Stochastic Complex Networks} \label{sec_obs}

Talking about complexity, it is noteworthy to distinguish between three stages of system structure: \textit{simple}, \textit{complicated}, and \textit{complex} \cite{Cotsaftis2009}. Simple systems are the building blocks for both complicated and complex systems \cite{Milo2002}. The difference between complicated and complex systems is due to the fact that in the latter, interactions between system components are fairly strong and somehow overshadow the component features. As a result, while a \textit{reductionist} approach may work for analyzing complicated systems, for complex systems, taking a \textit{holistic} approach is a must \cite{Cotsaftis2009}. In networks, moving from a sparse structure towards a dense structure can be interpreted as passing from a complicated network to a complex network.

When it comes to networks, in different branches of science and engineering, it is common to deal with sequential data gathered from the network. A large portion of our knowledge about a network, especially when it is large-scale and complex, cannot be presented in terms of quantities that can be measured directly. In such cases, building a model would be the logical basis for explaining the cause behind what we observe via the measurement process. This leads us to the notions of \textit{state} and \textit{state-space} model of a dynamic network, where the term ``dynamic'' may refer to time evolution of node state \cite{Liu2011_Controllability}, edge state \cite{Nepusz2012}, a combination of both, or even size and topology of the network \cite{setoodeh09}.

To investigate \textit{reconstructing} (i.e. \textit{estimating}) the state of dynamic networks from measuring the outputs of its monitor nodes, a key question is whether or not it is possible to do so using a given model of the dynamic network under study. This critical question that must be answered before choosing a proper estimation algorithm among different candidates, leads us to the concept of \textit{observability} \cite{Muske1997}. For deterministic networks, observability implies that an observer would be able to distinguish between different initial states based on measurements. In other words, an observer would be able to uniquely determine observable initial states from measurements \cite{Kailath1980}.

For defining observability in the context of networks, we may need a paradigm shift from the classic state trajectories to more abstract trajectory manifolds \cite{Cotsaftis2009}. To be more precise, in estimating the state, we may settle for finding a restricted initial subspace of the original state space instead of an individual initial state\cite{Liu2011_Reply}.

In \cite{Liu2013_Observability}, Liu, Slotine, and Barabasi proposed an intuitive method, which provides possible sets of necessary monitor nodes in a ``deterministic'' network, according to the Jacobian-based definition of observability. Additionally, they mentioned that any of the given sets may be sufficient in some specific cases. The method, which is called LSB hereafter, is based on a graph theory concept, known as strongly connected component (SCC). An SCC is a subgraph, in which there exists a directed path from each node to every other nodes. Although it is easy to implement the LSB algorithm, here are a few points that are worth thinking about:
\begin{itemize}
\item LSB results in a number of sets (called root SCCs), from each of which a node should be selected as one of the monitor nodes. However, LSB does not provide any further information about which of the nodes in each root SCC would be a better monitor node.
\item For most of dense and almost uniformly edge-distributed networks (e.g., Erd\H{o}s-R\'{e}nyi networks), which simply has one or a few SCCs, the LSB method hardly provides any practical clue about the monitor nodes. Table~\ref{tbl:table_LSB} elaborates on this critical problem.
\item  LSB is meant for deterministic networks, where the network model is completely known and the observation of monitor nodes is assumed to be perfect. It is mentioned in \cite{Liu2013_Observability} under the suggested future research topics that both assumptions may be violated in practice, where model uncertainties and measurement imperfections are involved. Indeed, our experiments demonstrate that for problems with modelling and measurement uncertainties, the practical monitor nodes may be different from what are suggested by LSB.
\end{itemize}
More importantly, in many practical cases, we are limited in the number of monitor nodes due to different reasons including limited computational resources. In a problem with limited number of monitor nodes, LSB provides no preference among the suggested monitor nodes and may therefore be used only as a clue for the selection of the monitor nodes.

\begin{table*}[th]
\caption{Number of monitor nodes based on the Liu-Slotine-Barabasi (LSB) method are compared for two basic network topologies of the same size, namely, scale-free and Erd\H{o}s-R\'{e}nyi (ER) random networks. Each row has roughly the same number of edges and the number of monitors are averaged over 1000 realizations and rounded up. For the scale-free networks, $\alpha$, $\beta$, and $\gamma$ are respectively the probabilities of adding a new node connected to an existing node chosen randomly according to the in-degree distribution, adding an edge between two existing nodes (one existing node is chosen randomly according to the in-degree distribution and the other is chosen randomly according to the out-degree distribution), and adding a new node connected to an existing node chosen randomly according to the out-degree distribution. Clearly, with the same number of nodes, the more dense the network is, the less number of monitor nodes is suggested by LSB. Similarly, LSB suggests considerably less number of necessary monitors for more uniformly-distributed networks. It is also noteworthy that for ER random networks, which are more dense than $5\%$, LSB provides almost no information about the monitor nodes. A similar problem happens for dense scale-free networks as well.}
\begin{center}
\begin{tabular}{|c|c|
>{\columncolor[HTML]{ECF4FF}}l |
>{\columncolor[HTML]{ECF4FF}}c ||
>{\columncolor[HTML]{FFFFC7}}c |
>{\columncolor[HTML]{FFFFC7}}c |}
\hline
 &  & \multicolumn{2}{c||}{\cellcolor[HTML]{BBDAFF}\textbf{Scale-free}} & \multicolumn{2}{c|}{\cellcolor[HTML]{FCFF2F}\textbf{ER Random}} \\ \cline{3-6}
\multirow{-2}{*}{Number of nodes} & \multirow{-2}{*}{\begin{tabular}[c]{@{}c@{}}Average number\\ of edges\end{tabular}} & \multicolumn{1}{c|}{\cellcolor[HTML]{ECF4FF}Parameters} & \begin{tabular}[c]{@{}c@{}}Avg. LSB monitors\\ ($\pm 1$)\end{tabular} & \begin{tabular}[c]{@{}c@{}}Probability for edge creation\\ $\in [0,1]$\end{tabular} & \begin{tabular}[c]{@{}c@{}}Avg. LSB monitors\\ ($\pm1$)\end{tabular} \\ \hline\hline
100 & 210 & \begin{tabular}[c]{@{}l@{}}$\alpha=0.41$\\ $\beta=0.54$\\ $\gamma=0.05$\end{tabular} & 74 & 0.021 & 12 \\ \hline
100 & 370 & \begin{tabular}[c]{@{}l@{}}$\alpha=0.21$\\ $\beta=0.74$\\ $\gamma=0.05$\end{tabular} & 67 & 0.037 & 3 \\ \hline
100 & 600 & \begin{tabular}[c]{@{}l@{}}$\alpha=0.41$\\ $\beta=0.54$\\ $\gamma=0.05$\end{tabular} & 56 & 0.060 & 1 \\ \hline
100 & 1620 & \begin{tabular}[c]{@{}l@{}}$\alpha=0.05$\\ $\beta=0.94$\\ $\gamma=0.01$\end{tabular} & 1 & 0.162 & 1 \\ \hline
\end{tabular}
\end{center}
\label{tbl:table_LSB}
\end{table*}

Going one step further to address real-world problems, the issue of stochasticity deserves special attention. In different applications, it is often desirable to predict next states based on collected data up to a certain time instant. Since the future is always uncertain, it is also preferred to have a measure that shows our confidence about the predictions; a probability distribution over possible future outcomes will do the job \cite{murphy02}.

For stochastic systems, there is not a unique definition of observability. However, most of the proposed definitions for observability of stochastic systems have roots in information theory. For instance, in \cite{Liu2011}, observability was defined on the basis of the concept of \textit{mutual information}:
 \begin{eqnarray}\label{eq:mutual_info}
  I(\X; \Y) = H(\X) - H(\X|\Y).
\end{eqnarray}
where $H(\X)$ denotes \textit{entropy} of $\X$ and $H(\X|\Y)$ is defined as the entropy of random variable $\X$ (i.e. state vector) conditional on the knowledge of random variable $\Y$ (i.e. measurement vector), hence the term \textit{conditional entropy}. According to \cite{Liu2011}, state $\X$ is unobservable from measurement $\Y$, if they are independent or equivalently $I(\X; \Y) = 0$; otherwise, $\X$ is observable from $\Y$. Since mutual information is nonnegative, equation (\ref{eq:mutual_info}) leads to the following conclusion: if either $H(\X) = 0$ or $H(\X|\Y) < H(\X)$, then $\X$ is observable. A deterministic system is either observable or unobservable but for stochastic systems, a \textit{degree of observability} can be defined, which varies between 0 and 1 \cite{Kam1987}.

Referring back to networks, in general, two sets of states can be considered for a network: \emph{physical states} and \emph{information states}, which are associated to physical dynamics and information dynamics, respectively \cite{Hero2011}. In \cite{Haykin2012_CC}, the notion of cognitive controller was proposed for controlling the information state as a counterpart to physical controller that controls the physical state. In the proposed framework, cognitive and physical controllers play complementary roles.

This paper proposes a systematic method for improving observability and therefore the quality of physical-state estimates in stochastic networks, based on the previously mentioned notion of stochastic observability. Cognitive controller will be able to increase the degree of observability, if
\begin{itemize}
\item the measure of information, which is chosen as the information state, is the entropy of the physical state (i.e. the entropic state), and
\item the role of cognitive controller is defined to minimize this entropy.
\end{itemize}
In this setup, a cognitive perceptor computes the mentioned entropic state and thereby sets the stage for cognitive control. As a result, the cognitive controller operates as the information supervisor of the network to address the previously mentioned issues of concern in a cycle-by-cycle manner. To be more precise, the cognitive controller algorithmically chooses the best set of monitor nodes from one cycle of perception-action to the next in a way to reduce the conditional entropy. In case of complete observability, the entropic state will approach zero \cite{Fatemi2014}; however, this is not the case in practice due to the ever presence of uncertainty and modelling imperfections.

\section{Complex Networks Viewed as the Environment of Cognitive Dynamic Systems} \label{sec_pac}

\begin{figure*}[t]
\centering
\includegraphics[trim = 0 -.2in 0 0, clip, width=4.6in]{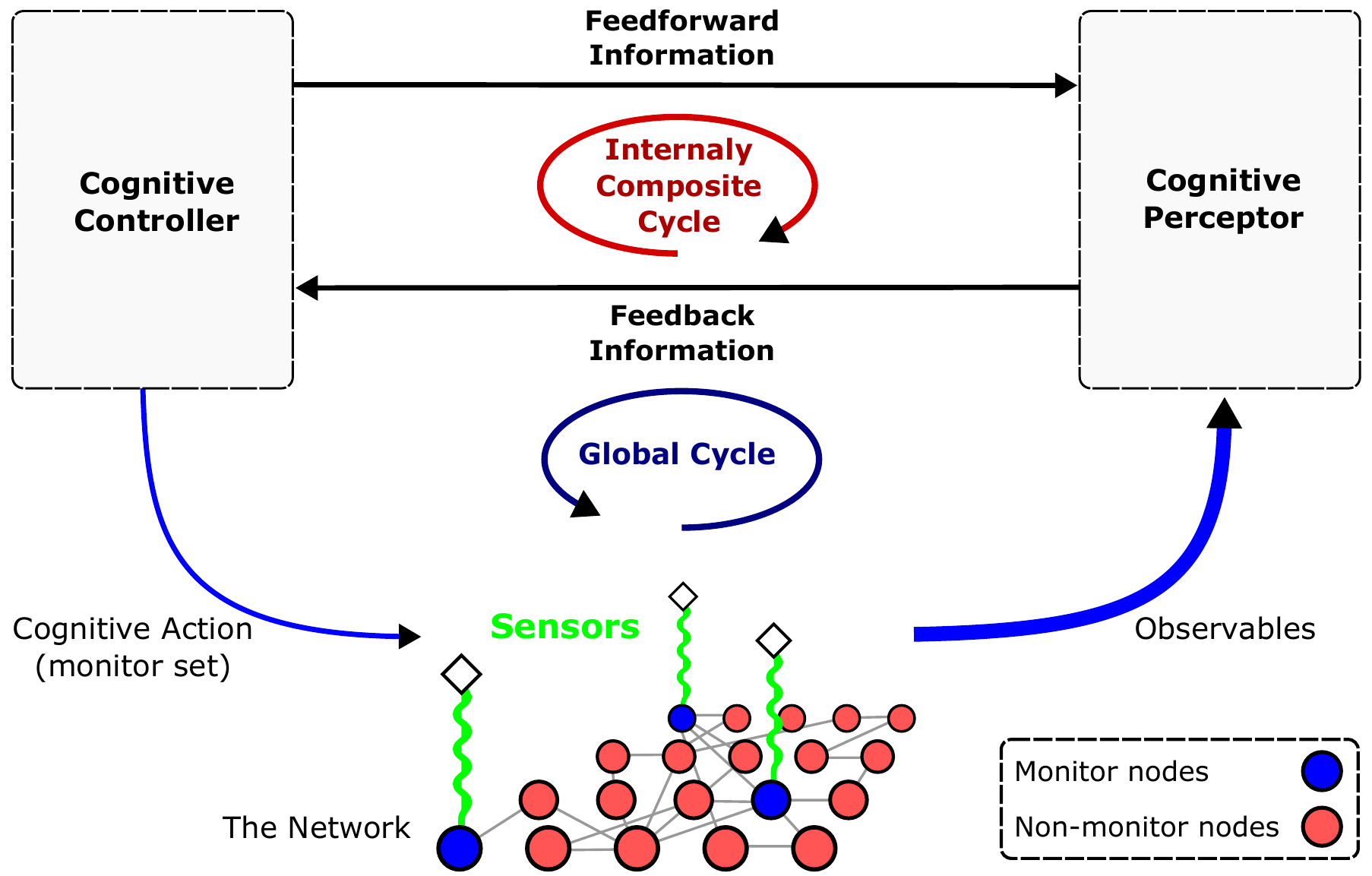}
\caption{Block diagram of the global perception-action cycle over a network, where a cognitive dynamic system acts as a supervisor. The nodes in blue are the monitor nodes and the diamonds are the observables.}
\label{fig_CDS}
\end{figure*}

Much has been written about the relationship between neuroscience and engineering. However, when it comes to cognitive neuroscience with emphasis on cognition, the Cognitive Dynamic System (CDS) first described in \cite{haykin2006} and later expanded in \cite{haykin2012}, is the closest description of such a system viewed from the perspective of Fuster's principles of cognition \cite{fuster_cortex_mind}. Fuster's principles are discussed in the Introduction; however, from the perspective of this article, it is the perception-action cycle that is the center of focus.

In the context of CDS, the environment is generic in terms of being an entity with any number of hidden states, which is seen only though the observables. As a result, it is quite natural to consider networks as the environment of a CDS with observables being the outputs of monitor nodes. In the structure depicted in Fig. \ref{fig_CDS}, the illustrated cognitive dynamic system indeed acts as a supervisor over a given network of interest in that it reconstructs the hidden state of the network on the sole basis of observing the monitor nodes. Furthermore, through the use of cognitive control, the cognitive dynamic system guarantees the accuracy of reconstructed state from each global cycle of perception-action to the next. In the following subsection, we first describe Bayesian perception of the network, which directly results in the definition of the so-called \emph{entropic state} that accounts for the mentioned information state. Then, the next two subsections discuss the cyclic directed information flow, which and the algorithmic processes involved in cognitive control.

\subsection{Bayesian Perception of Networks: The Two-state Model} \label{subsection_B}

We begin the global perception-action cycle by focusing on the perceptor on the right-hand side of Fig. \ref{fig_CDS}. The function of the perceptor is to monitor the network separately from the controller, and reconstruct the network state on the sole basis of extracting information from the observables. To be more specific, we may look to \emph{Bayesian filtering} \cite{Bayesian_filter_Ho} for estimating the hidden state of the network; using a state-space model (\ref{eq_state_space}) or (\ref{eq_state_space_nonlin}) that consists of a pair of equations: (a) \emph{process equation} that describes evolution of the state over time, which is contaminated by system noise, and (b) \emph{monitor equation}, which describes dependence of the incoming observables on the state of monitor nodes, corrupted by measurement noise. Optimal solution of the state estimation problem is given by the well-known \emph{Bayesian filter} \cite{Bayesian_filter_Ho}, at least in conceptual terms, which includes the special but important case of Kalman filter and its nonlinear versions \cite{kalman_1960, barshalom_estimation_book, oxford_handbook_nfiltering}. In a more general fashion, also for non-Gaussian environments, particle filters might be preferred to approximate the optimal Bayesian filter \cite{particle_filter_1, pf2004, Robert2005}. If we have a continuous-time process, then the Bayesian perceptor will take the form of a hybrid filter due to the fact that the observation process is still discrete in time.

As discussed in \cite{Fatemi2014}, the two-state model is an essential element in deriving the cognitive control algorithm. By definition, the two-state model embodies two distinct states, one of which is called the \emph{network state}, that is the vector of all the states attributed to the nodes (or edges or both) of the network. The second one is called the \emph{entropic state} of the perceptor, the source of which is attributed to the unavoidable presence of uncertainties in the environment as well as imperfections in the perceptor itself. These two states exactly corresponds to the ``physical'' and ``information'' states, which were previously discussed. Insofar as cognitive control is concerned, the two-state model is described in two steps as follows:
\begin{enumerate}
\item State-space model of the network, which is described by (\ref{eq_state_space}) or (\ref{eq_state_space_nonlin}).
\item Entropic state model of the perceptor, which is formally defined by the following equation:
\begin{align}
\mbox{Entropic-state equation: } ~~~H_{k} = \phi(p(\textbf{x}_{k}|\textbf{z}_{k}))
\end{align}
The $H_{k}$ is the entropic state at cycle $k$ in accordance with the state posterior $p(\textbf{x}_{k}|\textbf{z}_{k})$ in the Bayesian sense, which is computed in the perceptor\footnote{To emphasize the cycles, in which the state and the measurement are taken, in this paper, we may also use the notation $H_{k|k}$, in accordance with the subscripts in the posterior, $p(\textbf{x}_{k}|\textbf{z}_{k})$.}. As such, $H_{k}$ is the state of the perceptor and $\phi$ is a quantitative measure such as Shannon's entropy\footnote{Shannon's entropy for a random variable $X$, having the probability density function $p_{X}(x)$ in the sample space $\Omega$ is defined as\cite{cover_information}: $$H = \int_{\Omega} p_{X}(x) \log{\frac{1}{p_{X}(x)}} dx$$
Correspondingly, Shannon's entropy of network state $\textbf{x}_{k}$ with the posterior $p(\textbf{x}_{k}|\textbf{z}_{k})$ is defined as: $$H_{k} = \int_{\mathbb{R}^n} p(\textbf{x}_{k}|\textbf{z}_{k}) \log{\frac{1}{p(\textbf{x}_{k}|\textbf{z}_{k})}} d\textbf{x}_{k}$$ This entropy can be viewed as the perceptor state.}.
\end{enumerate}

It is important to note that in general, Shannon's entropy could assume the value zero; however, in cognitive control, the entropic state $H_{k}$ will always have a non-zero, positive value due to the fact that the environment always involves uncertainty and we can never reach perfect target-state reconstruction with $100\%$ accuracy.

By definition, the function of \emph{cognitive control} is to minimize the entropic state (i.e., state of the perceptor) on a cycle-by-cycle manner \cite{Haykin2012_CC}. Cognitive control therefore requires the entropic state, which is computed in the perceptor and then passed to the cognitive controller as \emph{feedback information}. Needless to say, this original definition of cognitive control matches the requirement of stochastic observability, as discussed previously.

The following subsection discusses the cyclic information flow and defines the cognitive controller as an optimal supervisor for the state reconstruction process by the perceptor.

\subsection{Cyclic Directed Information Flow} \label{subsection_C}

The \emph{global perception-action cycle}, depicted in Fig. \ref{fig_CDS}, plays a key role in a cognitive dynamic system; it is said to be global, in that it embodies the perceptor in the right-hand side of the figure, the cognitive controller in the left-hand side of the figure, and the monitored network, thereby constituting a closed-loop feedback system that includes the environment (i.e., the network in this context). The entropic state introduced in Subsection \ref{subsection_B} is indeed a measure of the lack of sufficient information for state-reconstruction in the perceptor. Next, the entropic state supplies the \emph{feedback information}, which is sent to the cognitive controller by the perceptor. With this feedback information at hand, the cognitive controller acts on the network, producing a change in the monitor nodes. Correspondingly, this change affects the amount of relevant information about the network, which is extracted from the new configuration of monitor nodes. A change is thereby produced in the feedback information and with it, a new action is taken on the network by the cognitive controller in the next perception-action cycle. These actions are called ``cognitive actions'' due to their role in controlling the directed information flow. To summarize, we may therefore define each cognitive action to be the selection of a possible set of monitor nodes. Continuing in this manner from one cycle of perception-action to the next, the cognitive dynamic system experiences a \emph{cyclic directed information flow}, as illustrated in Fig. \ref{fig_CDS}.

In addition to feedback information directed from the perceptor to the cognitive controller, there is also a \textit{feedforward information} link from the cognitive controller to the perceptor. In other words, the perceptor and the cognitive controller are reciprocally coupled. This important link provides the means for bypassing the network in order to ``predict'' the future global cycles for a hypothesized action. This feedforward link is the facilitator of \emph{predictive planning}.

\subsection{Summary of Cognitive Control}

The algorithmic steps involved in cognitive control from each cycle of perception-action to the next are summarized as follows (for further information, the reader is referred to \cite{Fatemi2014}):

\begin{enumerate}[A)]
\item \textbf{Initialization}:
\begin{enumerate}[i)]
\item Action Library: As described in Section \ref{sec_network}, for a network with $m$ accessible nodes and prescribed $q$ monitor nodes at each perception-action cycle, $\binom{m}{q}$ sets of monitor nodes will be available in total, the selection of which are considered to be cognitive actions in the action library of the CDS.
\item Value-to-go: To each cognitive action (set of monitor nodes), a value-to-go is allocated, which is initialized to zero.
\item Initial Action: One of the sets in the cognitive action library is then selected randomly at the very first cycle.
\end{enumerate}
\item \textbf{Cyclic Process}:
\begin{enumerate}[i)]
\item Given the observables, reconstruct the network state using Bayesian filtering and compute the state posterior through the well-known time-update and measurement update stages of filtering.
\item Compute the corresponding entropic state as the feedback information for cognitive control.
\item Compute the entropic reward and update the value-to-go function.
\item Compute the predictive planning updates using the internally composite cycle.
\item Repeat step ``iv'' for all hypothesized cognitive actions and lookahead predictions, as computationally permitted.
\item Using the resulting policy, select the best set of monitor nodes for the next cycle.
\end{enumerate}
\end{enumerate}

A direct consequence of using cognitive control is not only that it allows for the network structure to be dynamic, but it also results in finding an exact monitor set in each cycle as opposed to methods such as LSB, which only provide a collection of possible choices but not any exact choice.

Moreover, applying different constraints to monitor nodes are also permitted simply by defining the cognitive actions to be in accordance with the given constraints. This is another desirable feature of deploying cognitive control. The reason for this distinctive capability is that the cognitive controller finds the best cognitive action in the cognitive-action-space regardless of how this action-space has been defined. Therefore, we can define the cognitive-action-space in one form or another that best fits the design specifications of the problem at hand. For example, cognitive actions may be defined as sets of monitor nodes with prescribed cardinality or with inclusion/exclusion of a number of prescribed nodes. In the latter case, we may exclude some of network's nodes from being monitor nodes because they are \emph{inaccessible}. On the other hand, we may force the set of monitor nodes to include some prescribed nodes by simply defining the cognitive actions to be so.

Next section provides computer experiments in order to confirm the advantages of the proposed method and validate the claims made in the previous sections. For the sake of demonstrating the power of cognitive control, we explicitly restrict the cardinality of monitor sets to some prescribed values.

\section{Computational Experiments} \label{experiments}

In this section, we provide different examples to demonstrate the methodology just discussed. Our approach follows the one elaborated in \cite{Fatemi2014}. The first two sets of experiments pertain to the observability of linear networks. The third experiment will then examine the observability of a nonlinear benchmark process.

\subsection*{\textbf{Example 1:} A Small Linear Network}
\begin{figure}[t]
\centering
\includegraphics[trim = 0in 0in 0in 0in, clip, width=3.3in]{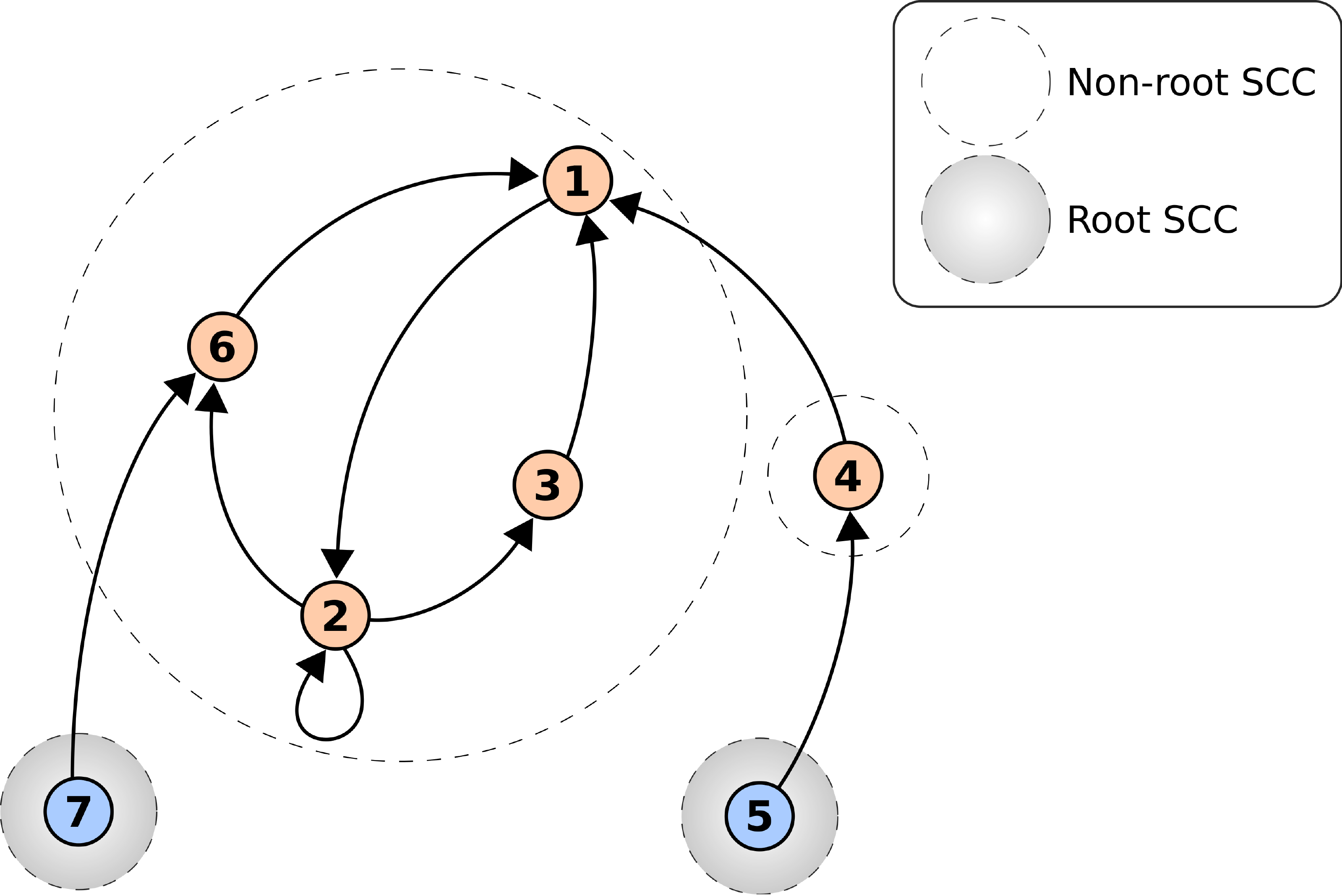}
\caption{Graphical illustration of the network in example 1. The numbered circles depict the seven nodes of the network. Dashed-line circles demonstrate strongly-connected components (SCC), where the shaded ones are the root SCC's that contain no inward edges. The nodes in blue (5 and 7) are the suggested monitor nodes by the LSB method.}
\label{fig_simple_graph}
\end{figure}
Consider a network of size $n$ (with $n$ number of nodes) with the adjacency matrix $\A$. Assume that all the nodes are accessible (i.e., $m=n$) but only $q<<n$ nodes are permitted to be monitored at each perception-action cycle. The main reason for this setup is that full monitoring of a complex network is not practically/computationally tractable. We demonstrate that a cognitive controller that minimizes the entropic state, is able to successfully select monitor nodes that minimize the state-reconstruction error of the network.

For the sake of demonstration of basic concepts, in this first example, we use a network of size $n=7$, with only one monitor node (i.e., $q=1$), and the following adjacency matrix:
\begin{align}
\nonumber \textbf{A} = \left[\begin{array}{ccccccc}0 & 0 & -0.3 & 0.9 & 0 & 0.4 & 0 \\1.2 & 1.2 & 0 & 0 & 0 & 0 & 0 \\0 & 0.4 & 0 & 0 & 0 & 0 & 0 \\0 & 0 & 0 & 0 & -0.5 & 0 & 0 \\0 & 0 & 0 & 0 & 0 & 0 & 0 \\0 & -0.6 & 0 & 0 & 0 & 0 & 1.7 \\0 & 0 & 0 & 0 & 0 & 0 & 0\end{array}\right]
\end{align}
As illustrated in Fig. \ref{fig_simple_graph}, the LSB method suggests nodes 5 and 7 as necessary monitors. Uncertainty in both state and monitor equations are modelled by additive zero-mean white Gaussian random processes. Under the Markovian assumption for state evolution, this problem therefore gives rise to the following state-space model:
\begin{align}
\nonumber \left\{\begin{array}{l}\textbf{x}_{k+1} = \textbf{A}^{T}\textbf{x}_{k} + \textbf{v}_{k} \\z_{k} = \textbf{e}_{j}\textbf{x}_{k} + w_{k}\end{array}\right.,
\end{align}
where, $\textbf{x}_{k}\in \mathbb{R}^{7}$ and $z_{k}\in \mathbb{R}$ are network's state and observation, respectively. Specifically, $\textbf{v}_{k}\sim\mathcal{N}(0,\textbf{Q})$ and $w_{k}\sim \mathcal{N}(0,\sigma_{w}^{2})$ are zero-mean, white Gaussian random processes with covariance matrix $\textbf{Q}$ for $\textbf{v}_{k}$ and variance $\sigma_{w}^{2}$ for $w_{k}$. Selection vector $\textbf{e}_{j}\in \mathbb{B}^{7}$, $\mathbb{B} = \{0,1\}$, is a row-vector with all of its elements equal to \textit{zero} except for the $j$-th element, which is \textit{one}. Hence, the $l$-th node of the network will be selected to be the monitor node if and only if $j=l$.  The initial state has been set to $1.00$ for all the states. Moreover, \textbf{Q} has been selected as the diagonal matrix with diagonal elements of $10^{-6}$, and $\sigma_{w}^{2} = 0.005$. For Gaussian processes similar to those involved in this problem, Shannon's entorpy can be shown to be $H_{k}=\frac{1}{2}\log ( \det \{(2\pi e)\textbf{P}_{k|k}\})$, with $\textbf{P}_{k|k}$ denoting the covariance of state reconstruction error vector at cycle $k$ given observation also at cycle $k$. We use $H_{k} = \mbox{trace}\{ \textbf{P}_{k|k} \} $ as the entropic state, where the \emph{trace} operator is sum of eigenvalues as opposed to \emph{determinant}, which is multiplication of eigenvalues. The reason is that \emph{trace} operator gives rise to the same result as Shanon's entropy due to the fact that eigenvalues of the positive-definite matrix $\textbf{P}_{k|k} $ are all positive values and logarithm is a monotonic function. However, \emph{trace} results in larger values for the entropic state, which is preferred for the sake of demonstration. The experiment run-time is 10 seconds with 10 observations per second.

\begin{figure*}[t]
\centering
\includegraphics[trim = .2in 0in .2in 0in, clip, width=4.7in]{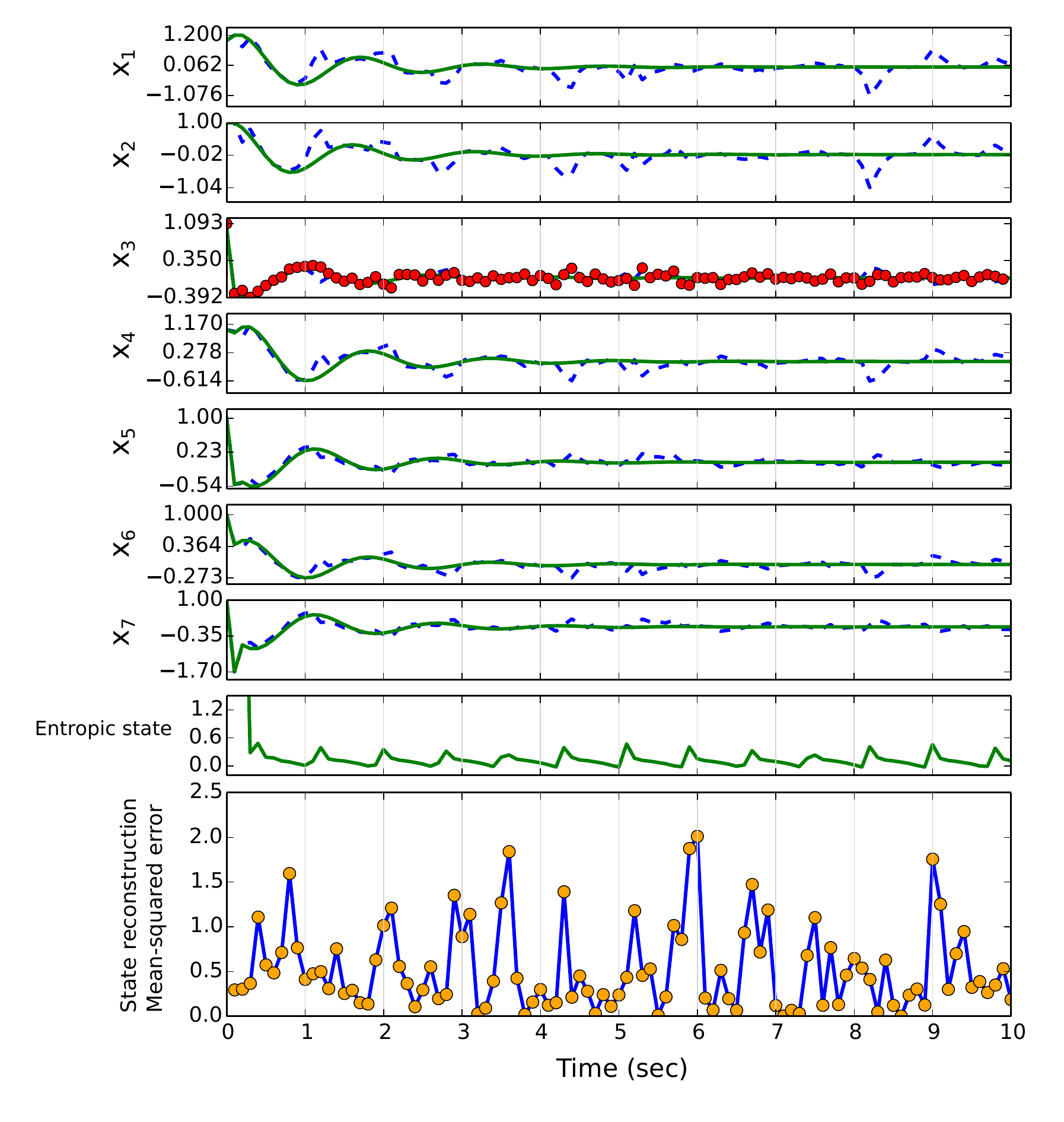}
\caption{Stochastic network observability (example 1): Out of seven states, only one state has been randomly selected to be monitored. Solid lines show the true states, dashed blue lines illustrate the estimated states resulting from the Kalman filter, and the red dots are noisy measurements, all coming from one randomly selected state ($x_{3}$ in this illustration). The entropic state only fluctuates over time with no convergence.}
\label{fig_netno}
\end{figure*}

The goal here is to find the best state to be monitored in each cycle using cognitive control. Using deterministic observability test, it is easy to show that deterministic version of this network (i.e., without noise terms) is not fully observable having a single monitor node. However, we may be able to minimize the entropic state by monitoring a ``proper'' node at each cycle with some minimum sample rate. In so doing, we produce a perception-action cycle by adopting a Kalman filter as the perceptor and the cognitive control algorithm as the controller to pick a proper monitor at each cycle. Fig. \ref{fig_netno} illustrates the problem \textit{without} cognitive controller. As expected, due to lack of observability, no correct estimation exists and both the entropic state and estimation error merely fluctuates with no actual convergence.

\begin{figure*}[t]
\centering
\includegraphics[trim = 0in 0in .2in 0in, clip, width=4.7in]{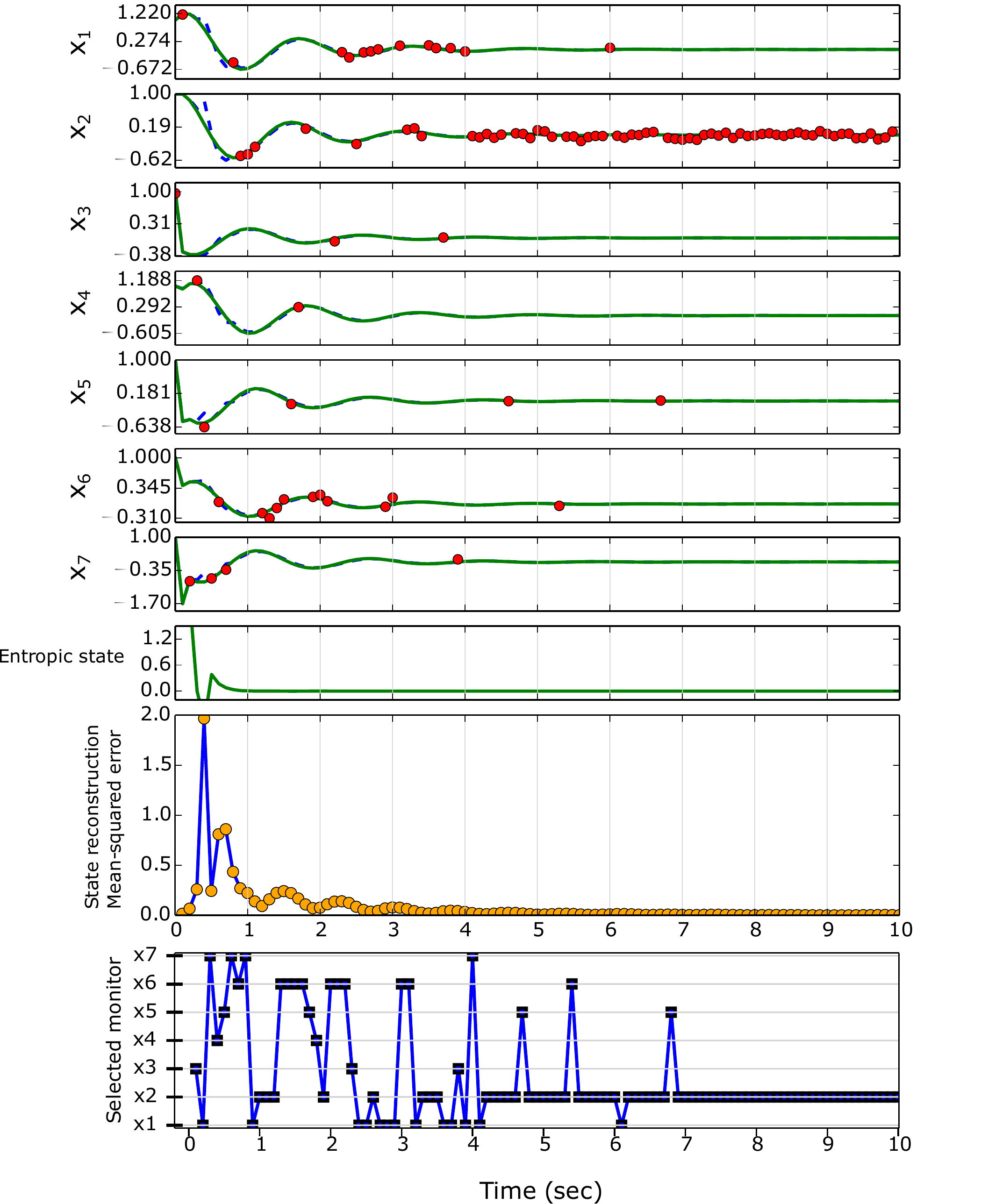}
\caption{Stochastic network observability with cognitive control for the same problem as in Fig. \ref{fig_netno} (example 1): As it can be seen, the entropic state becomes almost zero even only before the very first second of the experiment. It is noteworthy that the cognitive controller completely converges to a specific monitor node ($x_{2}$ in this case), which is different from the nodes suggested by the LSB method ($x_{5}$ and $x_{7}$). }
\label{fig_netwith1}
\end{figure*}

In Fig. \ref{fig_netwith1}, cognitive control has been deployed with 30 actions for planning in each global cycle and two predictive lookahead cycles for each of the hypothesized actions for planning. As illustrated, both the entropic state and the state reconstruction error have been minimized and stay close to zero. An interesting outcome is that the cognitive controller's action almost settles for choosing a specific node after the fourth second, and then completely stops switching after the seventh second.

\begin{figure*}[t]
\centering
\includegraphics[trim = 0in 0in 0in 2.25in, clip, width=4.6in]{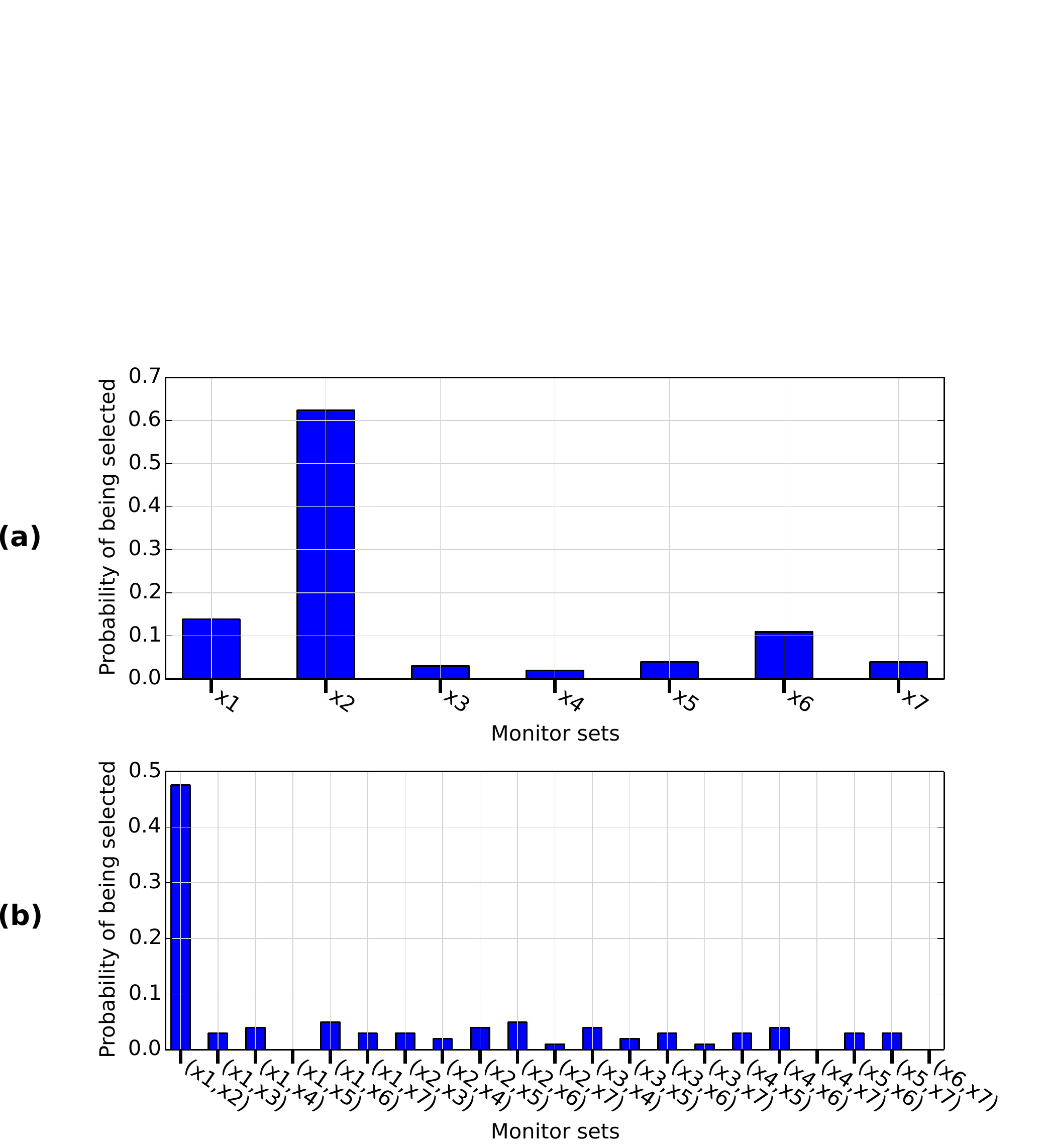}
\caption{Histogram of the selected nodes in example 1, using cognitive control with 30 random actions used for planning, each of which with two cycles of lookahead prediction into the future. Simulation results are averaged over 50 realizations. (a) Only one monitor node: It suggests $x_{2}$, $x_{1}$ and $x_{6}$ as the best monitors with the highest probabilities. (b) The number of monitor nodes is selected to be two: The histogram suggests the tuples ($x_{1}$, $x_{2}$), ($x_{1}$, $x_{6}$), and ($x_{2}$, $x_{6}$) as the best monitor sets with the highest probabilities. This is consistent with the case of only one monitor node.}
\label{fig_netwith_hist}
\end{figure*}

Next, we plot the histogram for monitor-node selection over the run time, which is averaged over 50 Monte Carlo realizations. The histogram, illustrated in Fig. \ref{fig_netwith_hist}-a, provides clue about which node proving itself to be more important in terms of network observability. Finally, we repeat the same experiment, this time with two monitor nodes, i.e., $q=2$. The resulting histogram of this experiment is depicted in Fig. \ref{fig_netwith_hist}-b, which nicely confirms the previous histogram.

\subsection*{\textbf{Example 2:} The Impact of Network's Topology on Observability of Complex Networks}

In the previous example, we discussed how cognitive control in a dynamic fashion picks the best monitor nodes, when the number of monitor nodes is limited. In this example, we expand on the methodology that explained in detail in Example 1, and implement cognitive control for two basic classes of complex networks, namely, Erd\H{o}s-R\'{e}nyi (ER) and scale-free random networks. Each class is examined with a number of different case-studies from sparse to dense networks. All the experiments involve uncertainty both in modelling as well as monitoring.

\begin{figure*}[t]
\centering
\includegraphics[trim = 0in 0in 0in 0in, clip, width=7.0in]{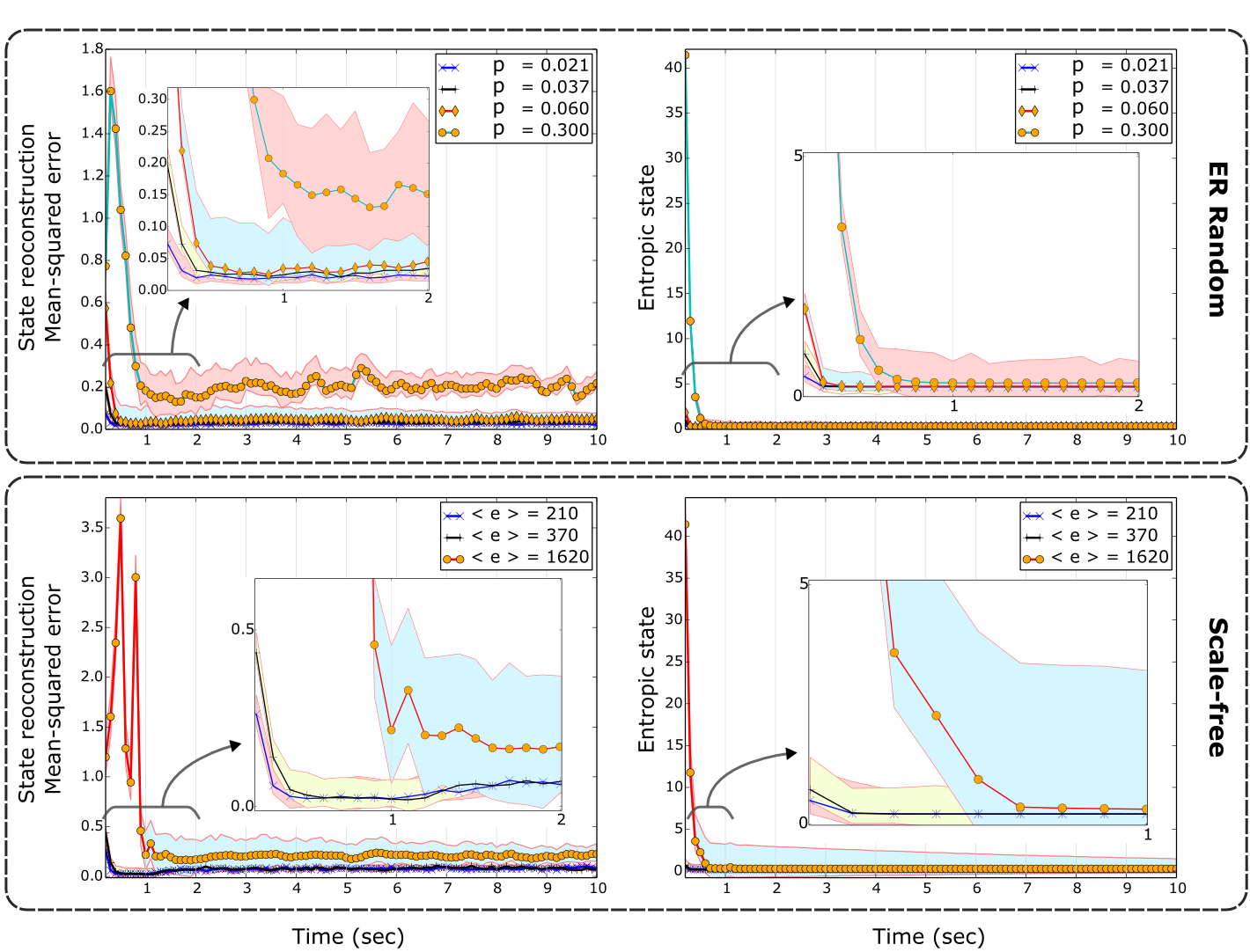}
\caption{Stochastic network observability in various configurations of Erd\H{o}s-R\'{e}nyi (ER) and scale-free networks. All the networks have 100 nodes and they are limited to have only one monitor. The parameters $p$ and $< e >$ denote probability for edge creation and average number of edges, respectively. For both types of networks, in addition to the entropic state, the state-reconstruction mean-squared error is also plotted to provide a measure for performance. In each global cycle, we use 150 hypothesized actions for planning and two predictive hypothesized cycles for each planning action. All the simulations are averaged over 50 realizations. For the sake of demonstration, close-ups of the confidence areas are also shown. For comparison, see also Table \ref{tbl:table_LSB}.}
\label{fig_netwith}
\end{figure*}

The results, illustrated in Fig. \ref{fig_netwith}, suggest that only one monitor node may still be sufficient to rapidly reach bounded and relatively small state-reconstruction error, provided that the monitor node changes dynamically over time (for a single static monitor, the Bayesian filter always crashes due to overflow of error covariance matrix). More importantly, the key result here is that in ER networks, which are uniform in terms of edge distribution, the state-reconstruction error is noticeably less (and more well-behaved) than their scale-free counterparts. This suggests that becoming more complex in terms of distance from a uniform structure may imply the need for more monitor nodes.

\textbf{Dense vs. Sparse Networks:} A basic question to be addressed is how \emph{edge} density and distribution affects the observability property of a network. Supported by Monte Carlo simulations depicted in Fig. \ref{fig_netwith}, our next key result is that as the network becomes more complex, both the entropic state and state-reconstruction error increase even in the presence of the cognitive controller. It implies the need for more monitor nodes as the network becomes more dense in terms of number of edges with fixed number of nodes. This result is clearly counterintuitive considering the hypothesis from LSB that the number of monitor nodes decreases as the network becomes denser and has less SCCs (see Table \ref{tbl:table_LSB}). Simply put, although the number of necessary monitor nodes suggested by LSB decreases, the actual number of sufficient monitor nodes will increase.

\subsection*{\textbf{Example 3:} A Benchmark Nonlinear Process}

In this last example, we follow a different approach: the role of cognitive control here is to dismiss the most redundant monitor(s) in each global cycle. In other words, we start with complete monitoring of all network nodes and we remove the node that is least informative in long-term (in dynamic programming sense). We then draw the histogram of the selected monitor nodes, which is not only over the entire run-time, but is also averaged over a large number of different realizations of the experiment. This approach provides a Monte Carlo based technique to find the best monitor nodes before the actual sensors are used in a lab setting, and may be very useful for practical sensor selection.

\begin{figure}[t]
\centering
\includegraphics[trim = 0in 0in 0in 0in, clip, width=3.4in]{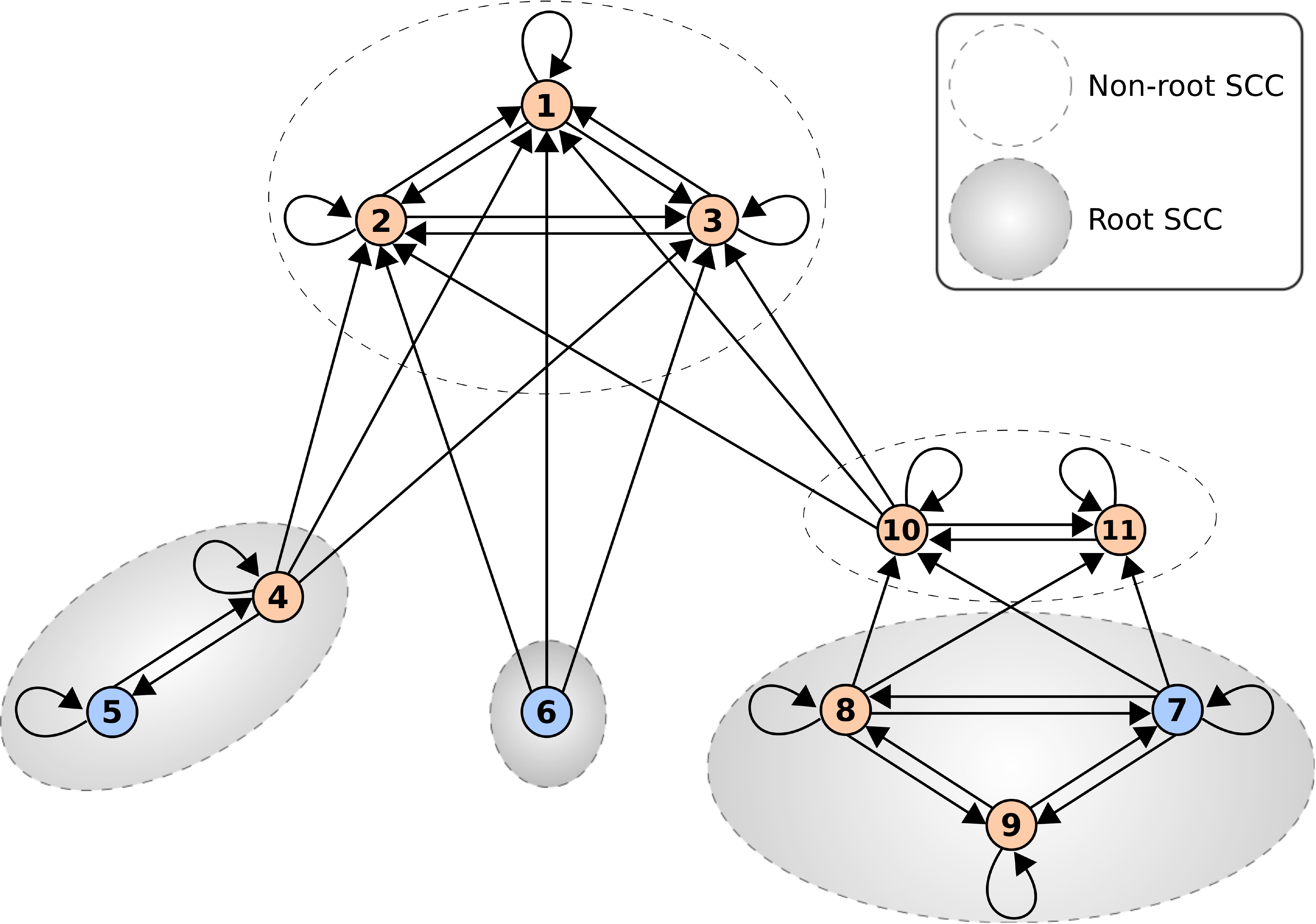}
\caption{Graphical illustration of the network in example 3. The numbered circles depict the nodes of the network. Dashed-line circles demonstrate strongly-connected components (SCC), where the shaded ones are the root SCC's that contain no inward edges. The ones in blue are the suggested monitor nodes by the LSB method.}
\label{fig_graph_nonlinear}
\end{figure}

For this example, we look into a benchmark nonlinear process, which is presented in \cite{Liu2013_Observability}, pertaining to a chemical reaction system with 11 species ($A$, $B$, $C$, ..., $J$, $K$) as depicted in Fig. \ref{fig_graph_nonlinear}. All the species are involved in the following four reactions:
\begin{align}
\nonumber &A + B + C \rightarrow D + F + J \\
\nonumber &D \leftrightarrow E \\
\nonumber &H + I \leftrightarrow G \\
\nonumber &J + K \rightarrow G + H
\end{align}
Because two of the reactions are reversible, we have six elementary reactions. Balance equations of the chemical reaction system are derived using the mass-action kinetics as the following \cite{Liu2013_Observability}:
\begin{align}
\nonumber \dot{x}_{1} &= -k_{1}x_{1}x_{2}x_{3} \\
\nonumber \dot{x}_{2} &= -k_{1}x_{1}x_{2}x_{3} \\
\nonumber \dot{x}_{3} &= -k_{1}x_{1}x_{2}x_{3} \\
\nonumber \dot{x}_{4} &= +k_{1}x_{1}x_{2}x_{3} - k_{2}x_{4} + k_{3} x_{5} \\
\nonumber \dot{x}_{5} &= +k_{2}x_{4} - k_{3} x_{5} \\
\nonumber \dot{x}_{6} &= +k_{1}x_{1}x_{2}x_{3} \\
\nonumber \dot{x}_{7} &= +k_{4}x_{8}x_{9} - k_{5}x_{7} + k_{6}x_{10}x_{11} \\
\nonumber \dot{x}_{8} &= -k_{4}x_{8}x_{9} + k_{5}x_{7} + k_{6}x_{10}x_{11} \\
\nonumber \dot{x}_{9} &= -k_{4}x_{8}x_{9} + k_{5}x_{7} \\
\nonumber \dot{x}_{10} &= +k_{1}x_{1}x_{2}x_{3} - k_{6}x_{10}x_{11} \\
\nonumber \dot{x}_{11} &= - k_{6}x_{10}x_{11}
\end{align}
where $x_{1}$, $x_{2}$, ..., $x_{11}$ denote concentrations of the 11 species, and rate constants of the six elementary reactions are given by $k_1$, $k_2$, ..., $k_6$, respectively. Based on the LSB method, the original deterministic system is suggested to have at least \emph{three} monitor nodes \cite{Liu2013_Observability}: $x_{6}$, one from $\{x_{4}, x_{5}\}$, and one from $\{x_{7}, x_{8}, x_{9}\}$.

We consider 1\% of uncertainty in both state and monitoring equations, presented by two white Gaussian random processes, respectively, which are mutually independent. Because the process equations are continuous-time and the monitoring process occurs in the form of digital sampling, which is discrete-time, we have to employ a \emph{hybrid} Bayesian filter \cite{barshalom_estimation_book}. Note also that the first derivatives with respect to most variables, the second derivatives with respect to \emph{all} variables, and all the higher order derivatives with respect to cross variables are all zero. Therefore, a hybrid extended Kalman filter (HEKF) will be the best choice \cite{barshalom_estimation_book}, since it will be very close to optimal. Nevertheless, it is important to say that because HEKF involves the use of a Range-Kutta ODE solver, its implementation involves additional approximations. Therefore, the result will not be on par with the linear discrete-time Kalman filter in the previous two examples. The experiment runs for 20 seconds with four sampling per second.

Using fixed monitors based on the LSB method, our simulations show that in 1000 Monte Carlo realizations of the experiment ALL have been crashed due to overflow of the estimation error covariance matrix. This fact implies that the information obtained from the monitor nodes suggested by the LSB method is considerably below sufficiency.

\begin{figure*}[t]
\centering
\includegraphics[trim = 0in -.2in 0in 0in, clip, width=7.0in]{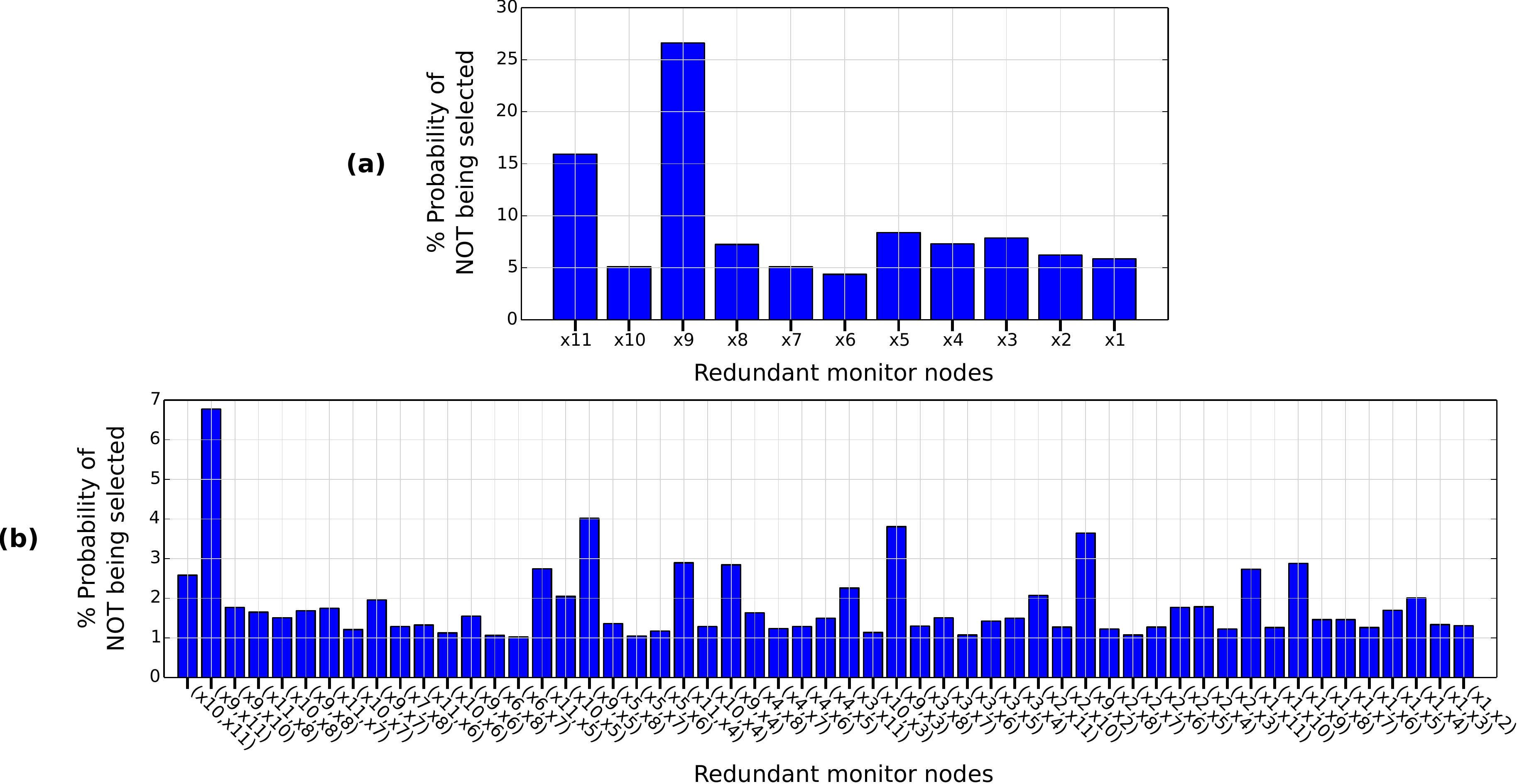}
\caption{Histogram of the redundant nodes in example 3, using cognitive control with 20 random actions used for planning, each of which with one cycle of lookahead prediction into the future. The simulations are averaged over 200 realizations. (a) Only one monitor node is considered as redundant: It suggests that $x_{9}$ and $x_{11}$ are the worst monitors with the highest probabilities. It also shows that $x_{6}$ is the most important one to be monitored, followed by $x_{7}$ and $x_{10}$ with the highest probabilities. The result is in partial agreement with the case of deterministic systems, (b) The number of redundant monitor nodes is selected to be two: The histogram suggests that the tuples that contain the nodes $x_{9}$, $x_{11}$, $x_{5}$, and $x_{3}$ are the worst monitor sets with the highest probabilities, which is in total agreement with the case of only one monitor node.}
\label{fig_nonlinear1}
\end{figure*}

Next, we deploy cognitive control to dynamically rank the best candidates to be monitored. To this end, we implement cognitive control in a way that it finds the worst monitor node in each global cycle. The cognitive controller uses  20 hypothesized action for planning in each global cycle and one predictive cycle for each of them. Fig. \ref{fig_nonlinear1} illustrates the resulting histogram. In the case that only one monitor node is considered as redundant, the histogram suggests that $x_{9}$ and $x_{11}$ are respectively the worst monitors with highest probability. It also shows that $x_{6}$ is the most important one to be monitored, followed by $x_{7}$ and $x_{10}$ with highest probability. The result is in partial agreement with the case of deterministic systems. We then repeat the experiment, but this time with dismissing the \emph{two} most redundant monitor nodes. In this case, the histogram suggests that the tuples that contain the nodes $x_{9}$, $x_{11}$, $x_{5}$, and $x_{3}$ respectively, are the worst monitor sets with highest probability, which is in total agreement with the case of only one redundant monitor node.

\section{Summary and Discussion} \label{sec_discussion}

In this paper, we studied the problem of observability in complex stochastic networks. The reported results demonstrated the fact that extending a good deterministic approach such as the LSB algorithm to stochastic networks is not straightforward because it may suggest an improper set of monitor nodes. Hence, we suggested to implement a cognitive dynamic system over the network of interest, for which the environment is the given network. Having the CDS, we will then be able to deploy cognitive control, which provides the best set of monitor nodes in a \emph{dynamic} manner. Regarding the proposed framework, the following points are noteworthy:

\begin{itemize}
\item A practical feature of deploying cognitive control is that in addition to optimizing design parameters such as the number of planning actions, ad hoc solutions can also be incorporated within the presented methodology in this paper so as to better match the dynamic monitoring process to the problem at hand. For example, the learning parameters for both learning and planning processes can be adaptively varied in the course of time. More interestingly, switching constraints may also be applied, if need be, to decrease switching between different monitor sets.

\item The proposed methodology may also be used offline using the Monte Carlo method. This way, the resulting histogram (similar to those presented in Example 3), can be used to provide realistic information about the best monitor nodes for the network of interest. This information may be very helpful for some real-world problems.

\item Our next key result based on the experiments is that as the network becomes more complex in terms of both edge density as well as distribution, the required number of monitor nodes increases, which is intuitively satisfying. This conclusive statement seems counterintuitive regarding the hypothesis favoured by the LSB method that decreasing the number of SCCs results in decreasing the number of necessary monitor nodes.

\item Regarding the potentials of CDS, it will be logically sound to claim that CDS may play a key role in the design of next generation of systems in future. The reasons include:
\begin{itemize}
\item Many, if not all, complex networks in real world involve uncertainty both in the modelling as well as in the monitoring stages. The CDS is by definition the paradigm that deals with uncertainty through perception and control of the directed flow of information in the best manner possible.

\item By means of cognitive control, information supervision is an intrinsic part of the CDS paradigm, which guarantees entropy reduction in the perception process in an optimal (or sub-optimal) manner.

\item The proposed methodology of this paper will benefit from future advancements of the CDS paradigm. Two possible ways to enrich this methodology are suggested as future research topics. An active perceptual memory can be incorporated to enhance the Bayesian estimation by feeding the filter with modelling parameters. Moreover, inclusion of the so-called \emph{pre-adaptive} mechanism \cite{Haykin2014} in the control side may help to cope with disturbances and intermittencies in the monitoring process.
\end{itemize}
\end{itemize}

\bibliographystyle{IEEEtran}
\bibliography{IEEEabrv,references_obs}

\balance

\end{document}